\documentstyle[epsf]{preprint}
\journalid{TSUJIMOTO et al.}{Chemical Evolution}{1999}
\articleid{1}{6}

\def\cm3{{\rm ~cm}^{-3}}
\def\ltsima{$\; \buildrel < \over \sim \;$}
\def\ltsim{\lower.5ex\hbox{\ltsima}}

\title{CHEMICAL EVOLUTION OF THE GALACTIC HALO THROUGH SUPERNOVA-INDUCED STAR 
FORMATION AND ITS IMPLICATION FOR POPULATION III STARS} {CHEMICAL
EVOLUTION OF THE GALACTIC HALO}
\author{Takuji Tsujimoto \\
{\em National
Astronomical Observatory, Mitaka-shi, Tokyo, 181-8588 Japan;} \\ 
{\em taku.tsujimoto@nao.ac.jp}, \\
Toshikazu Shigeyama$^{1}$, and Yuzuru Yoshii$^{1,\,2}$ \\
1) {\em Research Center for the Early Universe, Graduate School of 
Science, University of Tokyo,} \\ {\em Bunkyo-ku, Tokyo, 113-0033 Japan} \\
2){\em Institute of Astronomy, Faculty of Science, University of 
Tokyo, Mitaka-shi, Tokyo, 181-8588 Japan}}{T. Tsujimoto et al.}

\received{8 March 1999}
\accepted{30 April 1999}
\abstract{
A model for Galactic chemical evolution, driven by supernova-induced
star formation, is formulated and used to examine the nature of the
Galactic halo at early epochs.  In this model, new stars are formed
following each supernova event, thus their abundance pattern is
determined by the combination of heavy elements ejected from the
supernova itself and those elements which are already present in the
interstellar gas swept up by the supernova remnant.  The end result is
a prediction of large scatter in the abundance ratios among
low-metallicity stars, reflecting a different nucleosynthesis yield
for each Type II supernova with a different progenitor mass.
Formation of new stars is terminated when supernova remnants sweep up
too little gas to form shells. We show from calculations based on the
above scenario that (i) the observed [Fe/H] distribution for the
Galactic halo field stars can be reproduced without effectively
decreasing the heavy-element yields from Type II supernovae by some
manipulation required by previous models (e.g., via mass loss from the
early Galaxy, or later mixing with ``pristine'' hydrogen clouds), (ii)
the large observed scatter in the abundance ratio [Eu/Fe] for the most
metal-poor stars can also be reproduced, and (iii) the frequency
distribution of stars in the [Eu/Fe]$-$[Fe/H] plane can be predicted.
Our model suggests that the probability of identifying essentially
metal-free stars (Population III) in the local halo is around one in
$10^{3-4}$, provided that star formation in the halo is confined to
individual gas clouds with mass of $10^{6-7}\,M_\odot$ and that the
initial mass function of metal-free stars is not significantly
different from the Salpeter mass function.
}


\begin{document}
\section{INTRODUCTION}

In conventional chemical evolution models, stars are assumed to form
from well-mixed gas clouds at a rate proportional to some power of the
gas density (Schmidt 1959), thus inheriting the abundance pattern of
the gas at that time (e.g., Tinsley 1980).  This simplified treatment
of the complex star formation process has been remarkably successful
in accounting for the general features of the chemical compositions of
nearby stars and H$\;${\scriptsize II} regions in a consistent manner
(e.g., Pagel \& Patchett 1975; Matteucci \& Greggio 1986; Yoshii,
Tsujimoto, \& Nomoto 1996).  However, Shigeyama \& Tsujimoto (1998,
hereafter ST98) and Tsujimoto \& Shigeyama (1998, hereafter TS98)
argued that these conventional models cannot be applied to the early
stage of the Galactic halo, because observed abundance patterns of
extremely metal-deficient stars in the range of $-4<$[Fe/H]$<-2.5$
(McWilliam et al.~1995; Ryan, Norris, \& Beers 1996) are incompatible
with those predicted by these simple formulations. (see also Audouze
\& Silk 1995).

As an alternative, they proposed a scenario in which stars are born
from each supernova remnant (SNR), following a sweeping up of the
interstellar gas and formation of a dense shell immediately behind the
radiative shock front.  The dense shell is dynamically overstable
(Vishniac 1983; Ryu \& Vishniac 1987), so that it is broken into a few
thousand fragments, in each of which the self-gravity is unimportant.
According to Nakano (1998) it is likely that, for at least some of
these fragments, the ambient pressure outside the fragments exceeds
the critical value above which no equilibrium state exists.  Such
fragments will dynamically contract and eventually form stars.

This hypothesis sheds light on the entire process of early-epoch star
formation.  Galactic chemical evolution is envisioned as a successive
sequence of supernova (SN) explosions, shell formation, and resulting
star formation.  In this picture, the abundance pattern of each star
is set not only by the heavy elements ejected from the SN explosion
which directly preceeds local star formation, but also incorporates
those elements already present in the interstellar gas which are swept
up by the SNR.  The predicted stellar abundance patterns are thus
different from those of the gas at the time when stars form. This
difference can be quite large in the early stage of Galactic
evolution, when the metallicity in the gas is very low. The abundance
ratios of low-metallicity stars are predicted to exhibit a large
star-to-star scatter, depending in detail on the abundance patterns of
SN ejecta with different progenitor 
masses.

By incorporation of a sequence of SN-induced star formation into the
enrichment process of heavy elements, we describe the evolution of
stellar abundances in the Galactic halo characterized by the
inhomogeneous mixing of chemical compositions in dense shells and the
ambient medium.

\section{CHEMICAL EVOLUTION IN A CLOUD}

We assume that the star-forming process is confined to separate
clouds, of mass $M_c$, which make up the entire halo.  In this section
we present a formulation to describe the chemical evolution in a cloud
that is initially composed of metal-free Population III stars (Pop
III) and gas which has yet to form stars. The mass fraction of such
stars in our model is a parameter hereafter denoted by $x_{\rm III}$,
the massive ones of which eventually explode as Pop III SNe to
initiate chemical evolution.

All stars of subsequent generations are assumed to form in SNR shells.
The mass fraction of each shell which turns into stars is assumed to
be constant, and is denoted by $\epsilon$.  Heavy elements ejected
from an SN are assumed to be trapped and well-mixed within the SNR
shell.  Some of these elements go into stars of the next generation,
and the rest is left in the gas that will be mixed with the ambient
medium. The above process will repeat with increasing metallicity
until SNRs can no longer sweep up enough gas to form shells. No stars
will form from SNRs after this happens, and the process terminates.

When stars form as above, the star formation rate (SFR) 
${\dot M}_\ast(t)$ induced by Pop III SNe at time $t$ measured 
from the birth of their progenitor stars is given by 
\begin{equation}
{\dot M}_\ast (t)=\epsilon M_{\rm sh}({m_t {\rm ,}\, t}) {\phi 
({m_t})\over m_t}
\left| {dm_t\over dt} \right| x_{\rm III}M_c {\rm ,}
\end{equation}
and the SFR for later generations can be expressed as
\begin{equation}
{\dot M}_\ast (t)=\int_{\max(m_t,\, m_{{\rm SN}, l})}^{m_u}
\hspace{-1.5cm}dm \epsilon M_{\rm sh}({m {\rm ,}\, t}) 
{\phi(m)\over m }{\dot M}_\ast ({t-\tau (m)}){\rm ,}
\end{equation}
where $\tau(m)$ denotes the lifetime of a star of mass $m$, and $m_t$
is the stellar mass for which $\tau(m)=t$.  The initial mass function
(IMF) $\phi (m)$ used here is a Salpeter one, with a slope index of
$-1.35$. An upper mass limit of stars is assumed to be
$m_{u}=50\,M_\odot$; a lower mass limit of $m_{l}=0.05\,M_\odot$
(Tsujimoto et al.~1997) is also assumed. The lower mass limit for
stars that explode as SNe is taken to be $m_{{\rm SN}, l}=10\,
M_\odot$.  The mass of the shell $M_{\rm sh}(m,t)$ formed at time $t$
from an SN with progenitor mass $m$ is a sum of the mass of the SN
ejecta $M_{\rm ej}(m)$ and the mass of the swept-up gas $M_{\rm
sw}(m,t)$;
\begin{equation}
\label{eqn:masssh}
M_{\rm sh}({m{\rm ,}\, t})=M_{\rm ej}(m)+M_{\rm sw}({m{\rm ,}\,
t}){\rm ,}
\end{equation}
\noindent with
\begin{equation}
\label{eqn:masssw}
M_{\rm sw}({m{\rm ,}\, t})=\rho _{\rm g}(t){R^3 }
_{\rm SN}\left(E\left(m\right) ,\rho _{\rm g}(t)\right),
\end{equation}
where $\rho_{\rm g}(t)$ is the density of the interstellar gas and
$R_{\rm SN}\left(E(m),\rho_{\rm g}(t)\right)$ is the maximum radius of
the SNR shell.  $M_{\rm sw}(m,t)$ is insensitive to time $t$ because
$M_{\rm sw}$ is proportional to $\rho_{\rm g}(t)^{-0.06}$ (see ST98). 
The SN explosion energy is assumed to be $E(m)=10^{51}$ ergs,
irrespective of stellar mass $m$.  Thus a constant value of $M_{\rm
sw}(m,t) = 6.5\times10^4\,M_\odot$ will be 
used below.

Utilizing the SFR introduced in equations (1) and (2), the mass 
$M_{\rm g}(t)$ and the metallicity $Z_{\rm g}(t)$ of the gas will change 
with time according to the formulae:
\begin{equation}
 \label{eqn:chevg}
\hspace{-12pt}\displaystyle{dM_{\rm g} \over dt}=-{\dot M}_\ast
(t)+\int_{\max ({m_t{\rm ,}\, m_l})}^{m_{u}}\hspace{-1.2cm}
dmM_{\rm ej}(m){\phi
(m) \over m}{\dot M}_\ast ({t-\tau (m)}) {\rm ,}
\end{equation}
\noindent and
\begin{displaymath}
\displaystyle{d({z_{\rm g}M}_{\rm g}) \over dt}
=-\int_{\max ({m_t{\rm ,}\, m_{{\rm SN}, \,l}})}^{m_{u}}
\hspace{-1.8cm}dm
Z_\ast ({m {\rm ,}\, t})\epsilon M_{\rm sh}({m {\rm ,}\, t}) 
{\phi (m)\over m }
\end{displaymath}
\begin{displaymath}
\times{\dot M}_\ast ({t-\tau (m)})+\int_{\max 
({m_t{\rm ,}\, m_l})}^{m_{u}}\hspace{-1.2cm}
dm\left(M_{\rm ej}(m)-M_{\rm z}(m) \right){\phi (m)
\over m}
\end{displaymath}
\begin{displaymath}
\times\int_{\max ({m_{t-\tau (m)}{\rm ,}\, m_{{\rm SN},
\,l}})}^{m_u}\hspace{-2.2cm}
dm^\prime Z_\ast ({m^\prime
{\rm ,}\, t-\tau (m)})\epsilon M_{\rm sh}({m^\prime {\rm ,}\, t-\tau
(m)}){\phi ({m^\prime }) \over m^\prime }
\end{displaymath}
\begin{displaymath}
\times{\dot M}_\ast ({t-\tau(m)-\tau ({m^\prime })})
+\int_{\max ({m_t{\rm ,}\, m_l})}^{m_{u}}\hspace{-1.2cm}
dmM_{\rm z}(m){\phi (m)\over m}
\end{displaymath}
\begin{equation}
\label{eqn:chevm}
\hspace{-12pt}\times\int_{\max ({m_{t-\tau (m)}{\rm ,}\,
m_{{\rm SN}, l}})}^{m_{u}}\hspace{-2.7cm}
dm^\prime \epsilon M_{\rm sh}({m^\prime
{\rm ,}\, t-\tau(m)}){\phi ({m^\prime }) \over m^\prime }{\dot
M}_\ast({t-\tau(m)-\tau({m^\prime })}),
\end{equation}
where $M_{\rm z}(m)$ is the mass of all synthesized heavy elements 
ejected from a star with mass $m$, and $Z_\ast(m,\, t)$ is the metallicity 
of stars born at time $t$ from an SNR shell with progenitor mass $m$, which 
is defined as 
\begin{equation}
\label{eqn:massms}
Z_\ast ({m{\rm ,}\, t})={M_{\rm
z}(m)+Z_{\rm g}(t)M_{\rm sw}({m{\rm ,}\, t}) \over M_{\rm
ej}(m)+M_{\rm sw}({m{\rm ,}\, t})}. 
\end{equation}

Since $M_\ast(0)=x_{\rm III}M_c$ and $M_{\rm g}(0)=(1-x_{\rm III})M_c$
at $t=0$, after specifying $x_{\rm III}$ and $M_c$ and setting $Z_{\rm
g}(0)=0$, equations (5) and (6) are then
integrated with (1)-(4), and (7).
Star formation stops when the total mass involved in SNR shells
reaches the total mass of gas, i.e.,
\begin{equation}\label{eqn:term}
\hspace{-12pt}\int_{t-\Delta t}^t\hspace{-15pt}dt'
 \int_{\max({m_{t'}{\rm ,}\, m_{{\rm SN{\rm ,}}\,
l}})}^{m_{u}}\hspace{-60pt}dmM_{\rm sh}({m{\rm ,}\, t'})
{\phi (m) \over m}{\dot M}_\ast({t'-\tau ({m})})
= M_{\rm g}(t),
\end{equation}
where $\Delta t$ is the time for the turbulent motion in the ambient
medium to completely destroy the shape of an SNR shell.  When this
condition is satisfied, the cloud is composed of stars, hot tenuous
gas with negligible mass, and dense blobs of gas (an SNR shell would
be broken into blobs) which fill in a negligible volume of the cloud.
In the following calculation $\Delta t = 3\times 10^6$ yr is adopted
which corresponds to the crossing time of the turbulent flow over an
SNR.

The heavy-element yields from SNe are estimated directly from the
observed abundances in metal-poor stars, as opposed to the presently
uncertain predictions of theoretical supernovae models, using the
procedure proposed by ST98 and TS98. We note that the europium yield
is estimated from the data of Ryan et al.~(1996).

\section{RESULTS}
\begin{figure}[ht]
\begin{center}
\leavevmode
\epsfxsize=\columnwidth\epsfbox{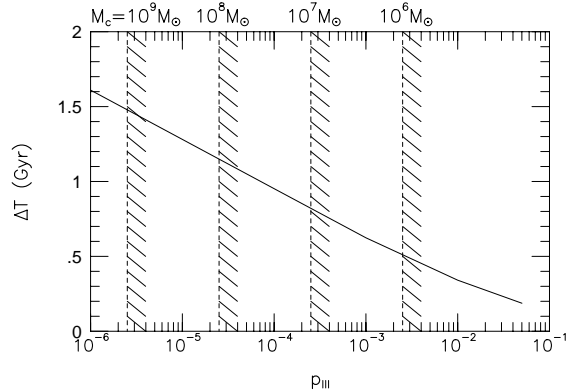}\hfil
\end{center}
\caption{The relation between the duration $\Delta T$ of star 
formation in a cloud and a probability $p_{\rm III}$ of observing Pop
III stars as a fraction of their number among all the long-lived stars
that have ever formed with $m<1 M_\odot$. The dashed lines denote the
strict lower bound of $p_{\rm III}$ for each mass $M_c$ of a cloud.}  
\end{figure}

\begin{figure}[ht]
\begin{center}
\leavevmode
\epsfxsize=\columnwidth\epsfbox{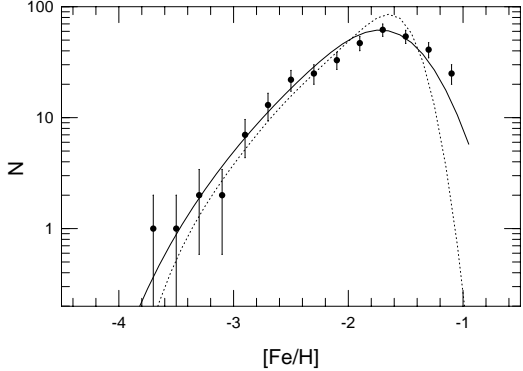}\hfil
\end{center}
\caption{The frequency distribution of the long-lived halo field stars 
for [Fe/H] $<-1$ against the iron abundance, compared with the
observations of Ryan \& Norris (1991). The observational data for
[Fe/H]$>-1$ are not shown because these data do not correspond to the
number pure halo stars due to an enormous contamination by disk stars
(see Ryan \& Norris 1991). The model curves have been convolved with
Gaussians having $\sigma$=0.3 dex (solid line) and $\sigma$=0.15 dex
(dotted line), respectively.} 
\end{figure}

\begin{figure}[ht]
\begin{center}
\leavevmode
\epsfxsize=\columnwidth\epsfbox{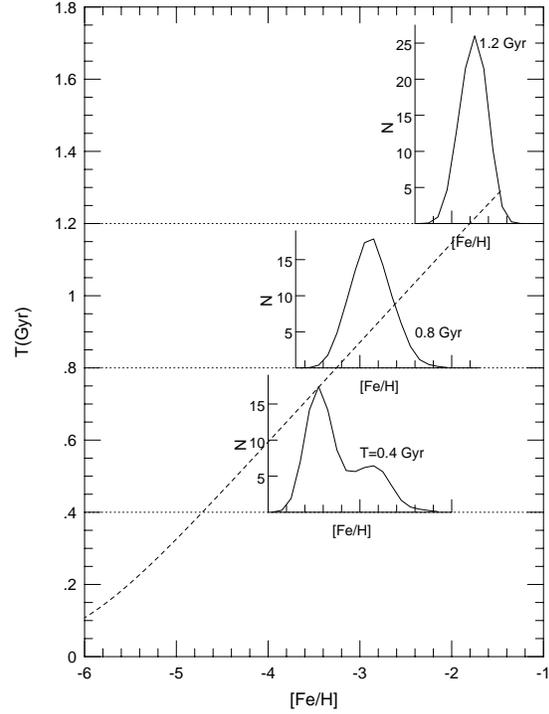}\hfil
\end{center}
\caption{The predicted [Fe/H] distribution functions of the 
long-lived stars at different ages of $4\times10^8, 8\times10^8,
1.2\times10^9$ yrs, in the age-[Fe/H] plane. The total number of stars
in each case is normalized to $100$.  The dashed line denotes the
age-metallicity relation of the gas.} 
\end{figure}

\begin{figure}[ht]
\begin{center}
\leavevmode
\epsfxsize=\columnwidth\epsfbox{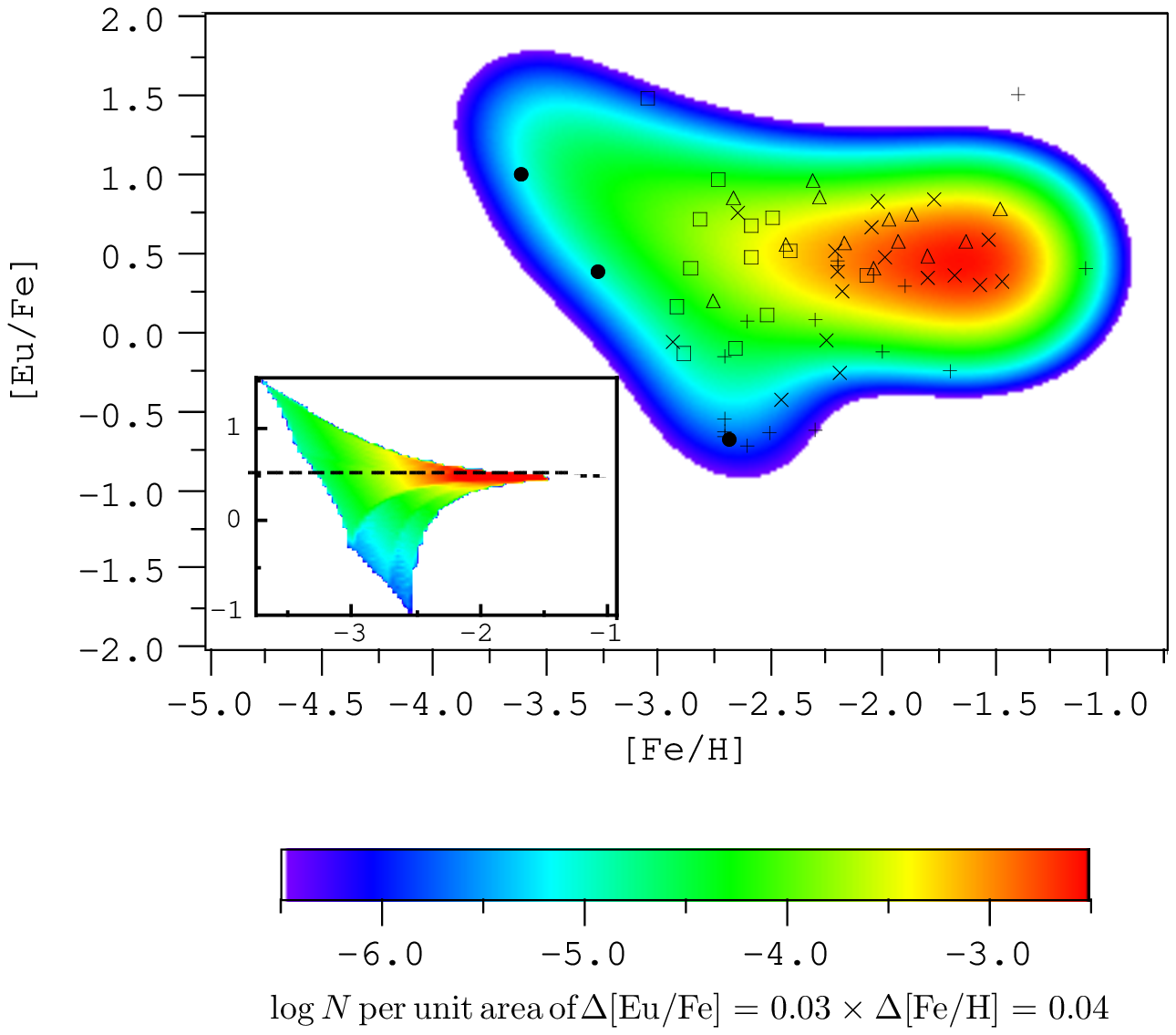}\hfil
\end{center}
\caption{The color-coded frequency distribution of the long-lived 
stars in the [Eu/Fe]$-$[Fe/H] plane convolved with a Gaussian having
$\sigma$=0.2 dex for [Eu/Fe] and $\sigma$=0.15 dex for [Fe/H], while
there is no convolution for the inset. The dashed line in the inset
denotes the [Eu/Fe]$-$[Fe/H] relation for the gas. The symbols
represent the data taken from various references (open squares:
McWilliam et al.~1995; filled squares: Ryan et al.~1996; plus signs:
Luck \& Bond 1985; open triangles: Gilroy et al.~1988; crosses: Magain
1989).} 
\end{figure}

From the formulation in the preceding section, we calculate the
chemical evolution for Galactic halo stars.  The free parameters in
our model are the mass fraction $x_{\rm III}$ of metal-free Pop III
stars initially formed in each cloud and the mass fraction $\epsilon$
of stars formed in a dense shell of each SNR.  The values of these
parameters are chosen to reproduce the observed [Fe/H] distribution
function of halo field stars for [Fe/H]$<-1$.  If $x_{\rm III}$ is
larger than about $1\times10^{-2}$, the total gas swept up by the
first SNRs exceeds the entire amount of available gas, and thus star
formation stops at the first or second generation.  In order for the
enrichment process to continue as a sequence of SN-induced star
formation, $x_{\rm III}$ must be less than about $1\times10^{-2}$.  In
addition, $\epsilon$ must be strictly confined within a narrow range
so that nearly one (slightly larger than one) massive star is born
from each SNR. This might imply that star formation is suppressed and
self-regulated by the photoionization from an OB star.  A value of
$\epsilon=4.3\times10^{-3}$ is found to give the best fit to the
observed [Fe/H] distribution function for various values of $x_{\rm
III} (<1\times10^{-2})$.  If $\epsilon$ slightly exceeds this critical
value, star formation will soon stop with little enrichment by heavy
elements.  On the other hand, if $\epsilon$ is taken slightly below
its critical value, star formation will proceed until most of the gas
is used up.  In such a scenario, the majority of halo stars become too
metal-rich to be consistent with observations.

The duration $\Delta T$ of star formation in a cloud is related to
$x_{\rm III}$ in the sense that a smaller $x_{\rm III}$ gives a longer
$\Delta T$.  Here we define a probability $p_{\rm III}$ of observing
Pop III stars as a fraction of their number among all of the
long-lived stars that have ever formed with $m<1 M_\odot$.  The
probability $p_{\rm III}$ decreases with increasing $\Delta T$ -- this
relation is shown by the thick line in Figure 1.  Each value of
$p_{\rm III}$ corresponds to about ten times $x_{\rm III}$, because
about 10 \% of $M_{\rm g}(0)$ has been converted to stars in the end.
We note that the most metal-poor halo stars exhibit abundance patterns
of genuine SN II origin, showing the overabundance of
$\alpha$-elements relative to iron Fe (e.g., Wheeler, Sneden, \&
Truran 1989), which means that they were born before SNe Ia start to
contaminate the gas with their products.  If the lower bound for the
lifetime of SN Ia is set at 1 Gyr (Yoshii, Tsujimoto, \& Kawara 1998),
the duration $\Delta T$ has an upper bound, and the probability
$p_{\rm III}$ should be larger than $10^{-4}$.  However, such an
argument does not hold if relatively few SNe Ia occur at [Fe/H]$<-1$,
as argued by Kobayashi et al.~(1998).

One absolute condition under which the SN-induced star formation
proceeds in a cloud is that at least one massive star of Pop III must
initially explode as an SN. Therefore, for each mass $M_c$ of a cloud
a lower bound of $p_{\rm III}$ exists, and is indicated in Figure 1 by
the vertical dashed line.  If $M_c$ is in a range of several $10^{6-7}
M_\odot$, which is inferred from the mass scale of the Galactic
globular clusters, one Pop III star might be observed in a complete
sample of $10^{3-4}$ halo stars.  This estimate depends critically on
the assumed IMF for Pop III stars.  For instance, if we adopt the
theoretical IMF for metal-free stars proposed by Yoshii \& Saio
(1986), the expected $p_{\rm III}$ is reduced by a factor of $15-45$.

Figure 2 shows the predicted stellar [Fe/H] distribution function for
$x_{\rm III}=10^{-6}$, as compared with the data obtained by Ryan \&
Norris (1991). The result has been convolved with a Gaussian of
$\sigma=0.15$ dex (dotted line) and $\sigma=0.3$ dex (solid line)
because measurement errors in [Fe/H] lie between these values (Ryan \&
Norris 1991).  Rather good agreement with the data is obtained for
[Fe/H]$<-1$ using the Salpeter IMF, without decreasing the
heavy-element yield by some manipulation such as mass loss from the
halo (Hartwick 1976; Searle
\& Zinn 1978; Bond 1981; Laird et al.~1988; Ryan \& Norris 1991). 
The termination of star formation at the gas metallicity of
[Fe/H]$\sim -1.5$ produces the peak of stellar frequency at
[Fe/H]$\sim -1.6$ as observed.  Only about 10\% of the initial gas has
been converted to halo stars, and the rest is left as the remaining
gas which may fall onto the disk to be consumed in formation of disk
stars.

Figure 3 shows the predicted [Fe/H] distribution function of stars
formed at a given age in the age$-$[Fe/H] plane. Three distribution
functions shown at different ages of $4\times10^8$, $8\times10^8$,
$1.2\times10^9$ yrs, normalized in such a way that the total number of
stars is 100.  The dashed line denotes the age-metallicity relation of
the gas.  It is important to note from this figure that the stellar
metallicity {\it does not} correspond to a unique age, but that the
scatter in stellar [Fe/H] among stars formed at a given age
progressively diminishes with time.  As can be appreciated from
inspection of the figure, some stars with abundance [Fe/H]$ > -3.0$
will form at essentially the same time as stars with abundances much
lower than this value.

Our primary concern is whether the observed stellar elemental
abundance patterns in the most metal-deficient halo stars can be
predicted by our model.  The abundance ratio [Eu/Fe] is a suitable
probe for this purpose, because this ratio in extremely
metal-deficient stars (McWilliam et al.~1995; Ryan et al.~1996) spans
over 2 dex, far exceeding the measurement errors.  Figure 4 shows the
color-coded frequency distribution of stars in the [Eu/Fe]$-$[Fe/H]
plane, normalized to unity when integrated over the entire area (see
the color bar for the scale).  In order to compare with the data, the
frequency distribution has been convolved with a Gaussian with
$\sigma=0.2$ dex for [Eu/Fe] and $\sigma=0.15$ dex for [Fe/H].  For
illustrative purposes the frequency distribution without convolution
is also shown in the inset.  The predicted [Eu/Fe]$-$[Fe/H] relation
for the gas is shown by the dashed line.  The predicted scatter in
[Eu/Fe] becomes smaller toward larger [Fe/H] and converges to a
plateau at $-2<$[Fe/H]$<-1$.  This tendency, which is also seen in
other elements, is understood by considering that the major
contribution to stellar metallicity has switched from the ejecta of
SNe to the metallicity in the interstellar gas swept up by SNRs (see
eq.[7]) in this abundance interval.

\section{CONCLUSION} 

We have presented a new model for Galactic chemical evolution based on
the SN-induced star formation hypothesis. 
An overview of our model
applied to the Galactic halo is as follows -- (1) The metal-free Pop
III stars form by some (as yet unspecified) mechanism in primordial
gas clouds of the Galactic halo.  (2) The most massive stars among
them explode as Type II SNe, which trigger a series of star formation
events.  (3) Star formation terminates when dense shells cannot be
formed from SNRs. (4) About 90 \% of the cloud mass remains unused in
star formation and may fall onto the still-forming Galactic disk.
Our model predicts that stellar abundance patterns are different from
those of the interstellar gas at the time when stars form.
Calculations presented here have shown that the observed features of a
large scatter in the europium to iron abundance ratios and the
metallicity distribution function with $\langle {\rm [Fe/H]}
\rangle\sim -1.6$ for halo field stars are naturally reproduced.  A
particularly attractive feature of our model is that agreement with
this mean stellar metallicity is achieved without decreasing the
value of heavy-element yields from SNe.

Our model strongly suggests the existence of metal-free Pop III stars,
because it is the first SN explosions among them that trigger a series
of SN-induced star formation.  Provided that star formation is
confined to separate clouds with mass of $10^{6-7} M_\odot$, and that
metal-free stars have a Salpeter IMF, it is estimated from the model
that one Pop III star would be found in a sample of $10^{3-4}$ halo
stars.  Ongoing surveys which seek to increase the numbers of stars
known at the lowest metal abundances (e.g., Beers, Preston, \&
Shectman 1992; Christlieb et al.~1999) 
provide the best hope for the eventual detection of this very {\it
first} generation of stars.  However, the probability of detecting
them would be smaller if the primordial IMF is enhanced on the massive
part compared to the Salpeter IMF (Yoshii \& Saio 1986) and/or if the
accretion of interstellar gas contaminates the surface of metal-free
stars (Yoshii 1981).  Both of these possible complications are
suitable for direct tests via high-resolution observations of a
sufficiently large sample of dwarf and giant stars with [Fe/H]$ <
-2.0$, which already exists (see Beers 1999 for a review).  For
an insight into the star formation process, it may be interesting
to consider that stars can be born from interstellar gas disturbed by
shock waves, implying that the first generation stars (Pop III stars)
could have been born in a dense shell formed by shock waves triggered
by a cloud-cloud collision in a protogalaxy.


\bigskip
This work has been partially supported by COE research (07CE2002) of
the Ministry of Education, Science, Culture, and Sports in Japan.
We would like to thank T. C. Beers for many fruitful discussions.


\begin{thebibliography}{}
\bibitem[]{}
Audouze, J., \& Silk, J. 1995, ApJ, 451, L49
\bibitem[]{}
Beers, T. C. 1999, in The Third Stromlo Symposium: The Galactic Halo,
eds. B. Gibson, T. Axelrod, \& M. Putman (Astronomical Society of the
Pacific; San Francisco), 165, 206
\bibitem[]{}
Beers, T. C., Preston, G. W., \& Shectman, S. A. 1992, AJ, 103, 1987
\bibitem[]{}
Bond, H. E. 1981, ApJ, 248, 606
\bibitem[]{}
Christlieb, N., Wisotzki, L., Reimers, D., Gehren, T., Reetz, J., \&
Beers, T. C. 1999, in The Third Stromlo Symposium: The Galactic Halo,
eds. B.  Gibson, T. Axelrod, \& M. Putman (Astronomical Society of the
Pacific; San Francisco), 165, 263
\bibitem[]{} 
Gilroy, K. K., Sneden, C., Pilachowski, C. A., \& Cowan,
J. J. 1988, {ApJ}, 327, 298
\bibitem[]{}
Hartwick, F. D. A. 1983, ApJ, 209, 418
\bibitem[]{}
Kobayashi, C., Tsujimoto, T., Nomoto, K., Hachisu, I., \& Kato, M. 1998, ApJ, 
503, L155
\bibitem[]{}
Laird, J. B., Rupen, M. P., Carney, B. W., \& Latham, D. W. 1988, AJ, 96, 
1908
\bibitem[]{}
Luck, R. E., \& Bond, H. E. 1985, {ApJ}, 292, 559
\bibitem[]{}
Magain, P. 1989, A\&A, 209, 211
\bibitem[]{}
Matteucci, F. \& Greggio, L. 1986, A\&A, 154, 279
\bibitem[McWilliam, Preston, Sneden, \& Searle 1995]{mcwilliam95}
McWilliam, A., Preston, G. W., Sneden, C., \& Searle, L. 1995, AJ, 109, 2757
\bibitem[]{}
Nakano, T. 1998, ApJ, 494, 587
\bibitem[]{}
Pagel, B. E. J., \& Patchett, B. E. 1975, MNRAS, 172, 13
\bibitem[Ryan, Norris, \& Beers 1996]{ryan96}
Ryan, S. G., Norris, J. E., \& Beers, T. C. 1996, ApJ, 471, 254
\bibitem[]{}
Ryan, S. G., \& Norris, J. E. 1991, AJ, 101, 1865
\bibitem[]{}
Ryu, D., \& Vishniac, E.T. 1987, ApJ, 313, 820
\bibitem[]{}
Schmidt, M. 1959, ApJ, 129, 243
\bibitem[]{}
Searle, L., \& Zinn, R. 1978, ApJ, 225, 357
\bibitem[Shigeyama \& Tsujimoto 1998]{shigeyama98}
Shigeyama, T., \& Tsujimoto, T. 1998, ApJ, 507, L135 (ST98)
\bibitem[]{}  
Tinsley, B. M. 1980, Fund Cosmic Phys., 5, 287
\bibitem[Tsujimoto \& Shigeyama 1998]{tsujimoto98}
Tsujimoto, T., \& Shigeyama, T. 1998, ApJ, 508,  L151 (TS98)
\bibitem[]{}
Tsujimoto, T., Yoshii, Y., Nomoto, K., Matteucci, F., Thielemann, F.-K., \& 
Hashimoto, M. 1997, ApJ, 483, 228
\bibitem[]{}
Vishniac, E.T. 1983, ApJ, 274, 152
\bibitem[]{}
Wheeler, J. C., Sneden, C., \& Truran, J. W. 1989, ARA\&A, 27, 279
\bibitem[]{}  
Yoshii, Y.  1981, A\&A, 97, 280
\bibitem[]{}  
Yoshii, Y., Tsujimoto, T., \& Kawara, K. 1998, ApJ, 507, L113
\bibitem[]{}  
Yoshii, Y., Tsujimoto, T., \& Nomoto, K. 1996, ApJ, 462, 266
\bibitem[]{}  
Yoshii, Y., \& Saio, H. 1986, ApJ, 301, 587
\end{thebibliography}
\end{document}